\documentstyle[12pt,epsfig]{article}  
\begin{document}
\noindent

\vspace*{-3cm}

\begin{center}
  \Large
{\bf Heavy Flavour Production at HERA}

    {T.Sloan}   \\
  \   Department of Physics, University of Lancaster, UK.\\
  \ (Talk at Moriond 2001)\\
\end{center}

\begin{abstract}
\noindent
Recent measurements of Charm and Beauty production at HERA are described.
The charm results are well described by NLO QCD calculations with 
a somewhat better description in the CCFM than in the DGLAP evolution 
schemes. However, the Beauty results in both photoproduction and in 
deep inelastic scattering (DIS) are poorly described by such 
calculations.
\end{abstract}



\section{Introduction}

Photon gluon fusion in leading order (LO) was first shown to be 
the mechanism governing charm production in deep inelastic scattering
(DIS) by the EMC \cite{EMC1,EMC2} at fixed target energies.  
At the higher energies at HERA higher order corrections 
become significant. In this paper the most recent 
HERA data on charm and beauty production in both DIS and photoproduction 
are compared to such computations at next to leading 
order (NLO). 

\section{Charm Production.}

New results on the differential cross sections for charm production 
have been announced by the H1 collaboration \cite{H11,H12}. The data 
are compared to the predictions of two Monte Carlo models which are 
based on NLO QCD. The HVQDIS \cite{HVQDIS} Monte Carlo model is based 
on the evolution at NLO of the momentum densities of the light quarks 
and the gluon of the proton in the DGLAP evolution scheme. The heavy 
quarks are then produced perturbatively by the 
photon gluon fusion mechanism. The CASCADE Monte Carlo uses a NLO 
calculation based on a solution of the CCFM equation \cite{CCFM}, 
which becomes BFKL-like as x$\rightarrow$0. It is found  
that the CASCADE model gives a better representation 
of the data than the HVQDIS model both in shape and normalisation.  
 

Both ZEUS and H1 have produced measurements of the 
charm contribution to the proton structure function $F_2$. The ZEUS 
collaboration has published measurements from inclusive D$^*$ 
production \cite{ZEUSD} and announced recently impressive
 measurements \cite{ZEUSe} from inclusive electron production 
where the electron comes from semileptonic charm decay. The 
latter measurements have extended the $Q^2$ range of the data up to 
565 GeV$^2$ and indicate that charm constitutes $\sim$40$\%$ of the 
proton structure function $F_2$ at this $Q^2$ and $x\sim$0.01. 

The H1 collaboration has announced its measurements from inclusive 
D$^*$ production \cite{H11,H12}. The values of $F_2^c$ were extracted 
using both the CASCADE and HVQDIS models to 
estimate the contribution from the unmeasurable region (about 70$\%$ 
of all the phase space). There are systematic 
shifts between the values using the different models to extrapolate 
the data into the full phase space \cite{H11,H12}. The data points 
extracted using HVQDIS tend to rise more steeply as $x$ decreases  
than those using 
CASCADE. The agreement of the both models and the data is reasonable 
but not perfect.
As with the differential cross sections the CASCADE model 
shows better agreement with the data than the HVQDIS model, each 
extracted within the framework of its own model. 

The data show that at these values of $x$ and $Q^2$ charm contributes 
a significant fraction to the inclusive structure function, $F_2$, 
of the proton. 
The data show a rapid rise with $Q^2$ at fixed $x$. In fact roughly 
half of the scaling violations seen in the inclusive structure 
function $F_2$ in the x range .0002 $ < x <$ .002 at the central 
$Q^2$ of the measurements can be attributed to charm production.  

\section{Beauty Production}

Both ZEUS and H1 have published measurements of the photoproduction 
($Q^2\sim$0) cross section for Beauty \cite{ZEUSb,H1bmu,H1bLi}. 
The ZEUS experiment \cite{ZEUSb} detects the electrons produced from 
semileptonic b decay to measure the cross section, obtaining the 
background from pair conversions from light quark fragmentation 
using the data.  The H1 experiment 
detects muons from semileptonic b decay \cite{H1bmu} measuring the 
background from muons from the decay and punch through from light 
hadrons from the data. In a second analysis 
\cite{H1bLi} the silicon vertex detector installed in H1 has been used 
to detect the b decays using the lifetime of the B meson which produces 
a finite impact parameter of the decay products at the primary vertex.
The lifetime method has also been applied by H1 to produce the first 
measurement of the b production cross section in DIS in the range 
$2 < Q^2 < 100$ GeV$^2$ \cite{H1bDIS}. The two H1 measured photoproduction  
cross sections are in good agreement but they lie significantly (by a factor 
$\sim$2) above the value of a QCD calculation at NLO \cite{FMNR}. 
The ZEUS photoproduction result is also a similar amount higher 
than the predictions of this model. The H1 DIS cross section is also 
found to be significantly higher than a NLO calculation based on the 
HVQDIS program \cite{HVQDIS}.


\section{Conclusions}

Charm production in DIS shows reasonable agreement with models 
based on NLO QCD. However, the agreement with the predictions 
of a Monte Carlo program based on the CCFM equation is better 
than with one based on DGLAP evolution. Perhaps this is an indication of the 
onset of BFKL behaviour. The measured Beauty cross sections are 
found to be larger than the predictions of NLO QCD.
It is very puzzling that charm production should be quite well reproduced 
by NLO QCD, top production at the TEVATRON is similarly well reproduced 
by NLO QCD \cite{TEVt} yet beauty production is not well reproduced by 
such calculations either at HERA or at the TEVATRON \cite{TEVb}. 
We heard at this conference an explanation for this phenomenon 
in which physics beyond the standard model is invoked \cite{ZackS}.

\section{Acknowledgements}
  
I thank all my colleagues in the H1 and ZEUS collaborations for their 
help in the preparation of this talk. 
I should also like to thank the organisers of the Conference for the 
lively and stimulating atmosphere which they have created.
  

%
\end{document}